# Synthesizing Audio from Silent Video using Sequence to Sequence Modeling

Hugo Garrido-Lestache Belinchon, Helina Mulugeta, Adam Haile*

Department of Electrical Engineering and Computer Science

Milwaukee School of Engineering

1025 N Broadway St, Milwaukee, WI 53202

{garrido-lestacheh, mulugetah, hailea} @msoe.edu

*Team Lead

# Contents

# Abstract

Generating audio from a video's visual context has multiple practical applications in improving how we interact with audio-visual media—for example, enhancing CCTV footage analysis, restoring historical videos (e.g., silent movies), and improving video generation models. We propose a novel method to generate audio from video using a sequence-to-sequence model, improving on prior work that used CNNs and WaveNet and faced sound diversity and generalization challenges. Our approach employs a 3D Vector Quantized Variational Autoencoder (VQ-VAE) to capture the video's spatial and temporal structures, decoding with a custom audio decoder for a broader range of sounds. Trained on the Youtube8M dataset segment, focusing on specific domains, our model aims to enhance applications like CCTV footage analysis, silent movie restoration, and video generation models.

# Introduction

Audio generation is a relatively recent field of AI that has yet to receive as much attention as video or image generation. This capability could prove valuable in many fields, such as enhancing CCTV footage analysis, generating sound effects for VFX, restoring historical videos, and improving video generation models. A field of audio generation that has been around for a long time and is more common is text-to-speech, which associates sounds with text. There has also been extensive research on the generation of music using AI. This research aims to enable the generation of arbitrary sounds, which are not necessarily speech or music.

This paper is a continuation of last year's research (Haile, Rozpadek, Mahmud, & Neuwirth, 2023). Using the same dataset, we suggest a new technique that would improve the efficiency and effectiveness of the model. The previous research encoded individual video frames and generated a relevant audio clip, and a custom WaveNet model was trained for video audio generation. WaveNet was called on each frame individually, which caused the audio generation process to be slow. It is limited in the generalization of audio it can generate as it is limited to one specific domain per weight. This research also attempted to manually categorize videos into one specific label. This pipeline generalized the video to a single tag, which caused issues when training as a single tag was associated with various audio outputs.

One way to generate audio, explored in this research's early stages, is using a large language model (LLM). This method is implemented by Google's VideoPoet (Kondratyuk, et al., 2023), "a language model capable of synthesizing high-quality video, with matching audio, from a large variety of conditioning signals." For training, the video is tokenized using MAGVIT-v2, and the audio is tokenized using SoundStream. Specialized tokens represent particular values, such as the beginning and end of the visual tokens or audio streams. The MAGVIT-v2 and SoundStream encoder and decoder pair would be used in the training phase to implement this for audio generation. In the inference stage, the SoundStream encoder would not be used, as the audio would not initially present. Just the same, the MAGVIT-v2 decoder would not be used.

The current model used in this research uses the VQ-VAE model. The training pipeline starts by splitting the video and audio. A VQ-VAE model encodes the video into a vector that can be used in the decoder. The Variational Autoencoder (VAE) converts the video into a stream that is then put into a Vector Quantizer (VQ) layer, which reduces it to a smaller set of numbers representing the video. This output is then fed into the decoder, which also takes in the audio input during training to associate the embeddings from the video to audio patterns. This trained decoder is then used in audio generation.

# Dataset

The Youtube8M dataset (Abu-El-Haija, et al., 2016), created by Google, is a manually annotated dataset of videos. It currently contains more than 6.1 million videos, with around 350k hours of video footage, and covers 3862 different classes. This dataset is becoming increasingly popular for video categorization due to its video diversity. Earlier studies have already used this dataset for related research (Haile, Rozpadek, Mahmud, & Neuwirth, 2023). For our study, we chose to focus on the "Airplane" class. This class was selected because airplanes typically have similar sounds, and the videos often have minimal background noise. This means that there was little data preprocessing required for the video and audio content. The Airplane category itself contains 35170 videos of various types. We hand-selected the videos used for training, as some videos in this category included content such as:

- Videos on paper airplanes or how to make them.
- Model airplane videos with more commentary than airplanes.
- News clips discussing airplanes but containing no visual content on airplanes themselves.

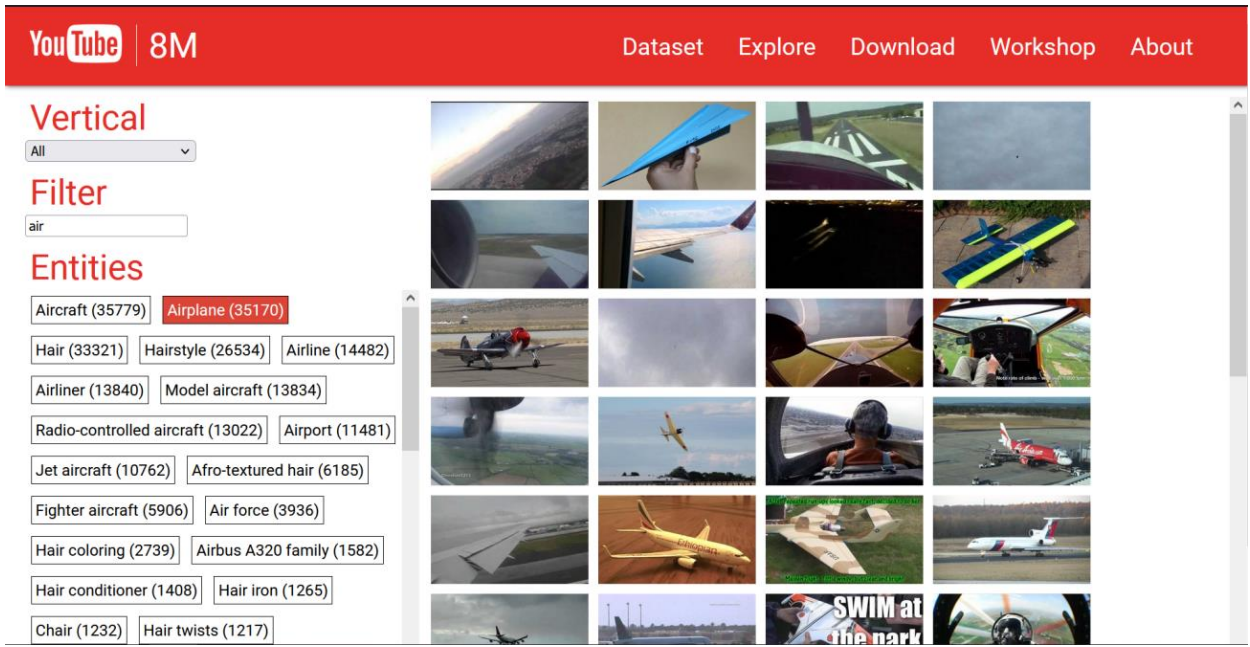

*Figure 1: A screenshot of the Youtube8M dashboard.*

## Feature Engineering

Data preprocessing for training has two steps. The first portion is for the encoder, which formats the videos to proper resolution and dimensionality. The second portion uses the encoder's outputs and the corresponding audio of the video to train the decoder.

For the encoder, the resolution of the video is scaled to 256 x 144. Color is maintained to preserve as much information as possible about the videos' content. Videos are split into 10-second segments, with the segments' audio spliced for the corresponding set of frames. Ten-second segments are then batched in groups of 2 at most for memory purposes.

The decoder is trained after the encoder. It utilizes both the encoder's discrete outputs and the segments of audio that were spliced separately. These audio segments are normalized to between -1 and 1 with a Tanh function. This allows us to have a ground truth of audio and means the model does not have to be highly complex.

On inference, only the video is processed. Any audio in the video is ignored. Similar preprocessing steps used in training the encoder are also applied to the inference video to ensure consistency. The only difference is that the video is not trimmed to 10 seconds but left at a given length.

## Model Overview

### Configuration

The entire model is an end-to-end encoder/decoder network, with the encoder model being a 3D Vector Quantized Variational Autoencoder (VQ-VAE) and the decoder being a fully connected neural network. The goal of the encoder is to create a discrete embedding of the video, while the decoder parses this discrete embedding to construct the audio waveform representation of the video. Outputs from the decoder are mathematically compared via Mean Squared Error (MSE) and manually compared to ensure quality.

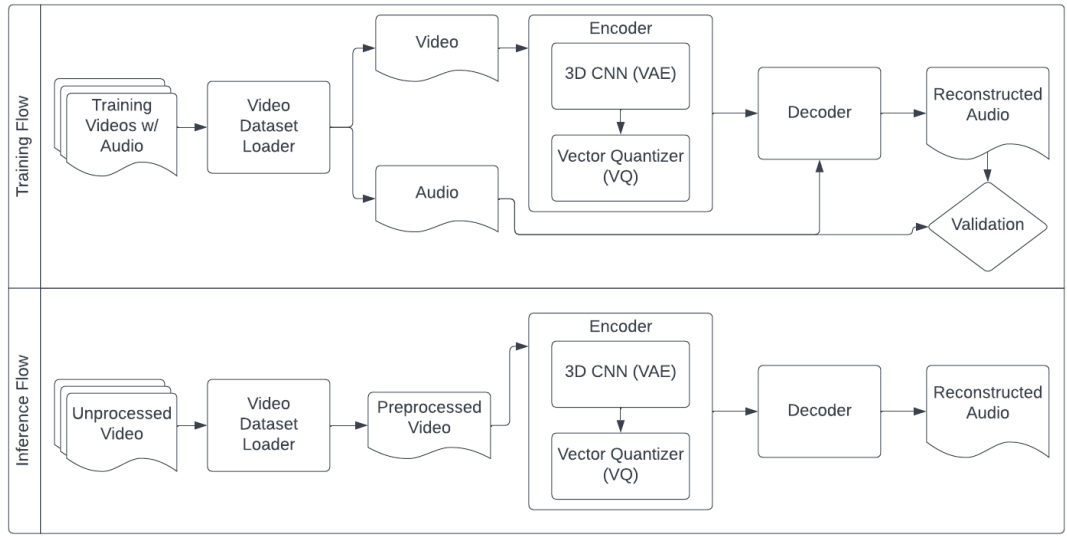

*Figure 2: Flow diagram of the model during both the training and inference stages. In inference, any provided audio is ignored as model cannot accept it as a modality.*

### Encoder - 3D Vector Quantized Variational Autoencoder

The 3D Vector Quantized Variational Autoencoder (VQ-VAE) comprises two repetitions of the 3D convolutional and ReLU layers before feeding into a Vector Quantizer. The VQ-VAE is used due to its speed and efficiency in training in comparison to a simple Variational Autoencoder (VAE) for decoding, while still being just as powerful as a VAE and avoiding key issues such as "Posterior collapse" (Oord, Vinyals, & Kavukcuoglu, 2018). In order to train the model, we prioritized the optimization of the encoder, as the decoder for our use would not be creating a reconstruction of the video but the audio representation. Our encoder learns how to map the input videos to a latent space representation, while the Vector Quantizer layer learns how to optimize its codebook of latent vectors.

The decoder layer is used to validate the VQ-VAE's results. This decoder reproduces the original video by using alternating transposed convolutions and ReLU layers with a final Sigmoid. The decoder is trained to compute the reconstruction loss.

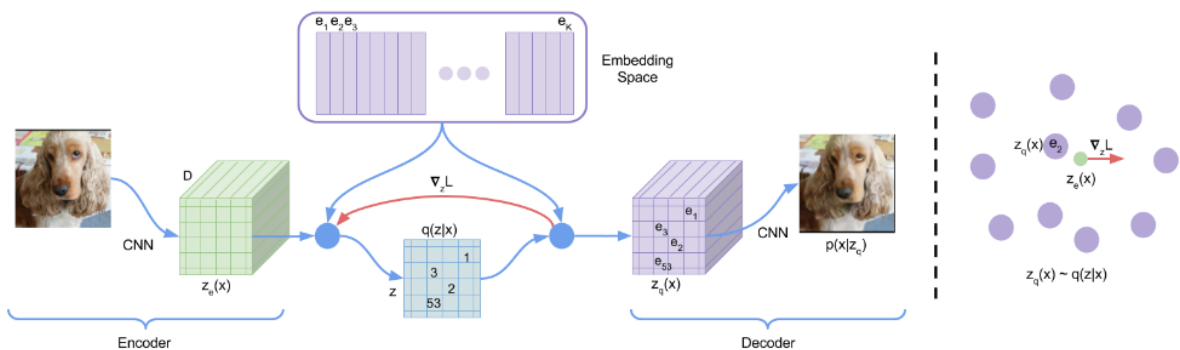

*Figure 3: (Oord, Vinyals, & Kavukcuoglu, 2018) Left: A figure describing the VQ-VAE. Right: Visualization of the embedding space. The output of the encoder z(x) is mapped to the nearest point e2. The gradient ∇z L (in red) will push the encoder to change its output, which could alter the configuration.*

During training, a forward pass creates a latent data representation. The Vector Quantizer then quantizes this latent representation. This quantization is then passed to the decoder to create the reconstruction. After the forward pass, we compute a backward pass, which returns the quantization loss and the commitment loss. The quantization loss moves the embeddings closer to the input encodings. The commitment loss ensures that the encoder commits to an embedding, preventing further training fluctuations. Then, the gradients are computed for the entire network. From here, we perform a gradient update to optimize the model's weights, including those in the VQ layer. The VQ also detaches the quantization before returning to ensure the VQ is not affected by backpropagation, resulting in a non-differentiable quantization operation. The inputs are attached to it instead, allowing the gradients to flow past the VQ when updating.

## Decoder – Feedforward Neural Network

The decoder is a series of fully connected linear layers with non-linear activations to transform the flattened encoded video input into an audio waveform. The encoded video is first transformed as a 3D tensor of batch size, number of embeddings, and the embedding dimension, into a 2D tensor. This 2D tensor contains the batch size and the flattened vector of all embeddings. For the first layer, the input size is kept aligned with the dimensionality of the flattened embedding space and the batch size. From here, the linear transformations and activation functions translate the data, and a final Tanh activation function is applied to normalize the data between -1 and 1. Tanh was used as the normalization function as it has been utilized in prior research with signal processing and reconstruction. (Wen, He, & Huang, 2022)

For training, first the audio is split from the video throughout our training data. Then the video is encoded into a latent representation and then passed through the decoder to generate an output.

From this the mean square error is calculated using the original sound and then perform gradient descent throughout the layers to adjust the weights and improve the output sound.

## Results

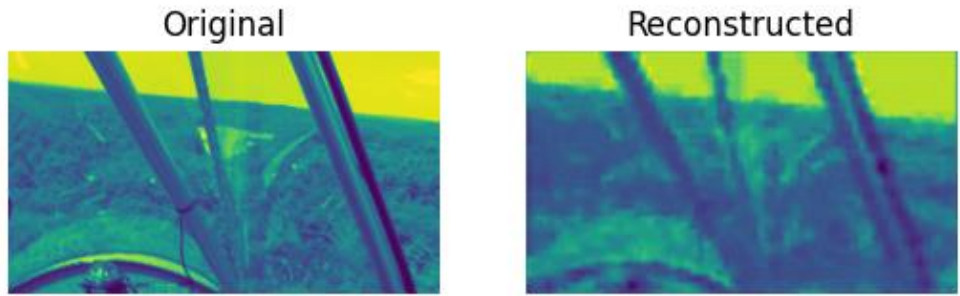

*Figure 4: A preprocessed frame of a testing video, along with the reconstruction result.*

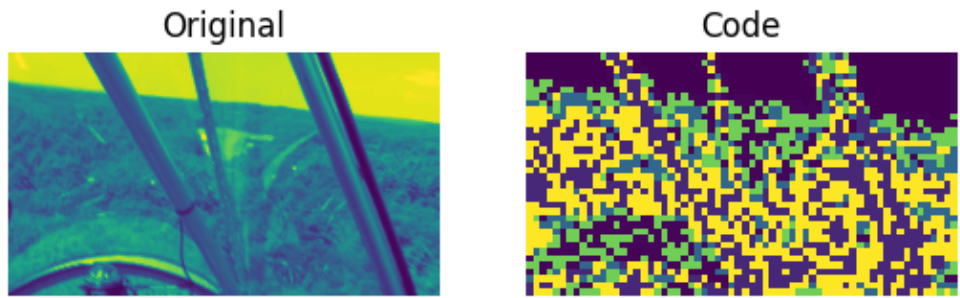

*Figure 5: A preprocessed frame of a testing video, and the associated discrete codes learned.*

For the VQ-VAE, we were able to train the model to successfully take a given video and develop a discrete representation of the video via frame-wise encoding. This discrete embedding is also able to be reversed and reconstruct a highly similar representation of the frame. These results map as well to videos and frames which the model has never seen before, as shown above.

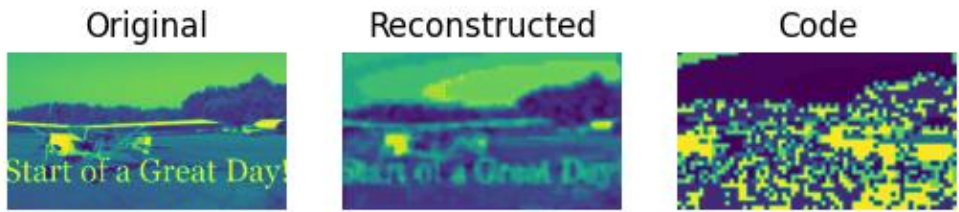

*Figure 6: A preprocessed frame of an airplane with some edited text overlaying the video, the reconstructed result of the airplane with a resemblance of that same text, and the discrete encoding of the same frame.*

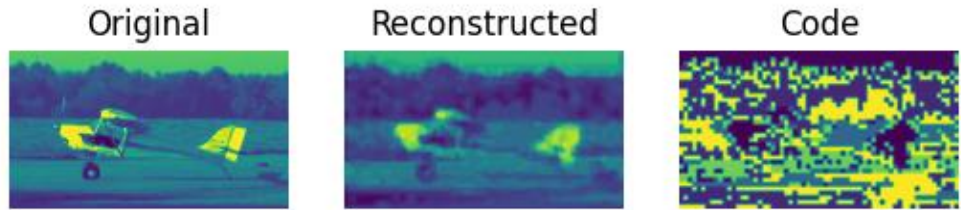

*Figure 7: A preprocessed frame of a small, single-engine airplane and the associated reconstruction showing the general shape of the same airplane, and the associated discrete encoding.*

## Limitations and Challenges

Throughout the work on this project many members dropped from the team due to other commitments. Due to this, the burden of work on the remaining team was more than doubled, and the complexity of the project given the timeframe increased greatly. As such, some compromises needed to be taken, more time was dedicated towards research, and much of this research is intended as theory with some experimentation. Final results were unachievable within the scope of this paper.

Additionally, this research was experimented utilizing only one T4 GPU. As such, much of the training process took a lot longer, and results were even more difficult to get when the model needed to be retrained. Ways this can be improved in the future would be to utilize multiple of MSOE ROSIEs GPU's. The ROSIE supercomputer, which was used to train the model, possesses a total of 20 T4 GPU nodes (Roughly 80 T4 GPUs) as well as 3 DGX Nodes (12 V100s). Utilizing the full power of ROSIE could provide a much-needed boost to the speed of development.

## Future Directions

In future iterations, there are many options we can take to further improve the model. One option is to spit the training to multiple GPUs via Horovod as this could allow for faster training of the model, allowing for more opportunities to tune the hyperparameters to reach the most capable version of this model.

Another idea would be to utilize an automated hyperparameter tuning process to achieve an optimal set of values for the model. A few implementations come from the Scikit-learn library, which has random and grid search algorithm options. Another such process is Bayesian Optimization, which iteratively creates a probability function based on the previous training results and sets hyperparameters to suggest the next best set of hyperparameters (Koehrsen, 2018).This process removes the guessing element from choosing hyperparameters and relies on trustworthy mathematical functions.

The best improvement for useability would be to train the model on a larger domain of videos other than just videos of planes. By expanding the domain of videos, the model should be capable of creating sound for a more extensive range of videos. Additionally, versions of the model could be developed for specific use cases. An example could be training the model on non-silent CCTV footage videos to generate the sound for silent versions of CCTV footage. This specialization can open the doors for many with disabilities that may prevent them from monitoring CCTV footage visually 24/7.

## Conclusion

In conclusion, this study explored the development and application of an end-to-end encoder/decoder network, utilizing a 3D Vector Quantized Variational Autoencoder (VQ-VAE) and a fully connected neural network, to synthesize audio from video content in the "Airplane" class of the Youtube8M dataset. Despite the project's limitations, such as limited time for experimentation and the constraints of balancing academic and work commitments, the research demonstrates potential pathways for advancing video-to-audio synthesis. The exploration into optimizing video processing and audio reconstruction reveals promising avenues for future work, including enhancing computational capabilities with distributed GPU learning, automated refinement/tuning of the hyperparameters, and broadening the domain of video content for model training. These directions could significantly improve the model's accuracy and applicability across various video categories and real-world scenarios. Ultimately, this project lays the groundwork for further innovations in multimedia content generation and offers insights into overcoming the challenges of synthesizing high-fidelity audio from video data.